\documentclass[10pt,superscriptaddress,twocolumn]{revtex4-2}

\usepackage{tikz}
\usepackage[cmex10]{amsmath}
\usepackage{amssymb}
\usepackage{physics}
\usepackage{upgreek}
\usepackage{titlesec}
\usepackage{siunitx}
\usepackage{here}
\usepackage{tabularx}
\usepackage{color}
\usepackage{comment}

\newcommand{\vp}{\varphi}
\newcommand{\ej}{E_\mathrm{J}}

\newcommand{\cj}{C_\mathrm{J}}

\usepackage{xcolor}
\usepackage{ulem}
\newcommand{\add}[1]{\textcolor{black}{#1}}
\newcommand{\erase}[1]{}

\usepackage{etoolbox}
\makeatletter
\renewcommand\thetable{\arabic{table}}            
\renewcommand\tablename{Table}                   
\patchcmd{\@caption}
  {\tablename\ \thetable.}
  {\tablename~\thetable:}
  {}{}
\makeatother

\usepackage{hyperref}
\hypersetup{
    colorlinks=true,
    linktocpage=true,
    linkcolor=blue,
    filecolor=magenta,      
    urlcolor=cyan,
    bookmarks=true,}
    
\graphicspath{{image2/}}

\begin{document}

\title{\add{Spectral Properties of Two Superconducting Artificial Atoms Coupled to a Resonator in the Ultrastrong Coupling Regime}}

\author{A. Tomonaga}
\email{akiyoshi.tomonaga@aist.go.jp}
\thanks{Current affiliation: National Institute of Advanced Industrial Science and Technology (AIST), Tsukuba, Ibaraki 305-8563, Japan}
\affiliation{Department of Physics, Tokyo University of Science, Shinjuku, Tokyo 162--0825, Japan}
\affiliation{Center for Quantum Computing, RIKEN, Wakoshi, Saitama 351--0198, Japan}

\author{R. Stassi}
\affiliation{Center for Quantum Computing, RIKEN, Wakoshi, Saitama 351--0198, Japan}
\affiliation{Dipartimento di Scienze Matematiche e Informatiche, Scienze Fisiche e Scienze della Terra, Universit\`a di Messina, I-98166 Messina, Italy}

\author{H. Mukai}
\affiliation{Department of Physics, Tokyo University of Science, Shinjuku, Tokyo 162--0825, Japan}
\affiliation{Center for Quantum Computing, RIKEN, Wakoshi, Saitama 351--0198, Japan}

\author{F. Nori}
\affiliation{Center for Quantum Computing, RIKEN, Wakoshi, Saitama 351--0198, Japan}
\affiliation{Physics Department, The University of Michigan, Ann Arbor, Michigan 48109-1040, USA.}

\author{F. Yoshihara}
\affiliation{Department of Physics, Tokyo University of Science, Shinjuku, Tokyo 162--0825, Japan}
\affiliation{Advanced ICT Research Institute, National Institute of Information and Communications Technology, Koganei, Tokyo 184-8795, Japan}

\author{J. S. Tsai}
\email{tsai@riken.jp}
\affiliation{Department of Physics, Tokyo University of Science, Shinjuku, Tokyo 162--0825, Japan}
\affiliation{Center for Quantum Computing, RIKEN, Wakoshi, Saitama 351--0198, Japan}

\begin{abstract}
We experimentally investigate a superconducting circuit composed of two flux qubits ultrastrongly coupled to a common $LC$ resonator. Owing to the large anharmonicity of the flux qubits, the system can be described well by a generalized Dicke Hamiltonian containing spin--spin interaction terms.
\add{In the experimentally measured spectrum, we observed two key phenomena. First}, an avoided level crossing provides evidence of the exotic interaction that allows the simultaneous excitation of two artificial atoms by absorbing one photon from the resonator. \add{Second, we identified a pronounced spectral asymmetry that is a clear signature of light–matter decoupling.} This multi-atom ultrastrongly coupled system opens the door to studying novel processes for quantum optics and quantum-information tasks on a chip.
\end{abstract}

\maketitle

% \begingroup
% \renewcommand{\thefootnote}{\fnsymbol{footnote}}
% \footnotetext[1]{e-mail: akiyoshi.tomonaga@aist.go.jp, tsai@riken.jp ${}^{7}$ Current affiliation: National Institute of Advanced Industrial Science and Technology (AIST), Tsukuba, Ibaraki 305-8563, Japan}
% \endgroup

%%%%%%%%%%%%%%%%%%%%%%%%%%%%%%%%%%%%%%%%%%%

\section*{Introduction}
Superconducting circuits provide a versatile and flexible platform for modeling various quantum systems~\cite{nakamura_coherent_1999,gu_microwave_2017,krantz_quantum_2019,kjaergaard_superconducting_2020,blais_circuit_2021,kwon_gate-based_2021}. 
In this platform, artificial atoms can be designed to have tailored energy transitions and controllable interactions with microwave photons~\cite{gu_microwave_2017}. Moreover, superconducting circuits have also become one of the main platforms for scalable quantum information processing and quantum simulation~\cite{krantz_quantum_2019,blais_circuit_2021,gu_microwave_2017,kjaergaard_superconducting_2020,kwon_gate-based_2021}.

Taking advantage of the high electromagnetic field in a one-dimensional resonator and the huge dipole moment of artificial atoms, these systems achieve a stronger light--matter interaction than those at bare atomic or resonator frequencies~\cite{niemczyk_circuit_2010,yoshihara_superconducting_2017,forn-diaz_ultrastrong_2017,bosman_multi-mode_2017,ao_extremely_2023,kockum_ultrastrong_2019,forn-diaz_ultrastrong_2019}.
This ultrastrong (deep-strong) interaction might lead to promising applications, such as high-speed and high-efficiency quantum information processing devices~\cite{romero_ultrafast_2012,kyaw_creation_2015,Wang2016Holonomic,wang_ultrafast_2017,stassi_scalable_2020,chen_fast_2021}. 
In this coupling regime, several unique physical phenomena have been predicted, and now, some of these are realized experimentally.

Important theoretical predictions \erase{indicate}\add{include}, for example, \erase{the possibility of observing} quantum vacuum radiation and entanglement from the ground state~\cite{de2007quantum,PhysRevA.81.042311,stassi_spontaneous_2013,cirio2016ground,stassi2023unveiling}\add{; multi-excitation exchanges between qubits and resonators \cite{garziano2015multiphoton}; or physical processes analogous to parametric down-conversion \cite{stassi2017quantum,Kockum_Deterministic}}.
In 2016, a theoretical work showed that one photon can simultaneously excite two atoms~\cite{garziano_one_2016,ball_two_2016}.
This effect \erase{is}\add{should be} observable if the atoms are ultrastrongly coupled with a cavity mode and the parity symmetry of the atoms is broken.
Similar to Rabi oscillations, this is a coherent and unitary process where the atoms can jointly absorb or emit one photon~\cite{garziano_one_2016,Kockum_Deterministic}.

\erase{Among the experimental works realized in the ultrastrong coupling regime, the observation of the induced parity symmetry breaking of an ancillary artificial atom is worth mentioning~[31].}
% \add{Among the experimental works in the ultrastrong coupling regime, it is worth mentioning the observation of: (i) the induced parity symmetry breaking of an ancillary artificial atom ~\cite{wang2023probing}; and  (ii) the multi-excitation exchange between one flux qubit and a waveguide resonator~\cite{niemczyk_circuit_2010}. In the first }
\erase{In this work, it was shown that, in an atom--light system \erase{that is }in the ultrastrong coupling regime, when the parity symmetry is broken, the light field acquires a coherence in the ground state that induces symmetry breaking in an ancillary flux qubit weakly coupled to the same field~[30]. }
% \add{The second work~\cite{niemczyk_circuit_2010} is the first realization of the ultrastrong coupling regime in circuit QED. As evidence of this achievement, the authors present an avoided level crossing that indicates the exchange of excitations between the qubit and two cavity modes.} 
\add{The first realization of the ultrastrong coupling regime in circuit QED~\cite{niemczyk_circuit_2010} was evidenced by an avoided level crossing indicating the exchange of excitations between the qubit and two cavity modes; namely, the multi-excitation exchange between one flux qubit and a waveguide resonator. Later, it was shown~\cite{garziano2014vacuum,wang2023probing} that, when the parity symmetry is broken in an atom–light system in the ultrastrong coupling regime, the light field acquires a coherence in the ground state that induces symmetry breaking in an ancillary flux qubit weakly coupled to the same field.}

Here, we experimentally investigate a circuit composed of two flux qubits ultrastrongly coupled to a common $LC$ resonator. Flux qubits, which form artificial atoms, share the same inductor with the $LC$ resonator, consequently they interact with each other. This system is described by the Dicke Hamiltonian generalized to include atomic longitudinal couplings and spin--spin interaction terms.

In the measured spectrum, away from the flux qubit optimal point (where the parity symmetry of the system is broken), we observe an energy-level anticrossing, which indicates hybridization between the states $\ket{gg1}$ and $\ket{ee0}$, where $\ket{g}$($\ket{e}$) and $\ket{0}$ respectively indicate the atomic ground (excited) and zero photon states. This is the fingerprint of the interaction that allows one photon to simultaneously excite two atoms, as well as the reverse process.
Here, when the system is set up in the one-photon state, the artificial atoms and the resonator can exchange excitations exhibiting Rabi-like oscillations.

Since the atom--light and atom--atom interactions are very strong, the system states should be strongly hybridized, and the clear observation of the ``one--photon--exciting--two--atoms'' effect would not be straightforward. 
However, by studying the generalized Dicke Hamiltonian, we found (i) a partial suppression of the transverse interaction given by a competition between the spin--spin interaction and the spins--light interaction and (ii) the ``decoupling'' of the longitudinal interaction that depends on the sign of the external flux bias. \add{The decoupling justifies the asymmetry in the measured spectrum.}

\add{While a multimode cavity coupled to a single atom, as in~\cite{niemczyk_circuit_2010}, allows frequency-selective interactions, its photon conversion capability is limited by mode spacing. In contrast, multiple atoms coupled to a single cavity (as in our work) can facilitate the conversion of a single photon into multiple lower-frequency excitations. Despite superficial similarities, these two circuits are different in configuration and serve distinct roles.}

% \color{blue}
% On the other hand, a recent experiment~\cite{mehta_down-conversion_2023} on many-body localization (MBL), not studied in our work, shows that the down-conversion of a single photon into multiple photons is suppressed due to localization in Fock space. This is in stark contrast to our system, which is designed to actively promote both down- and up-conversion processes.

\add{
On the other hand, a recent experiment~\cite{mehta_down-conversion_2023} based on many-body localization (MBL), not studied in our work, shows the suppression of the photon down-conversion, due to MBL.
This is in stark contrast to our system, which is designed to actively promote both down- and up-conversion processes. 
Moreover, Ref.~\onlinecite{mehta_down-conversion_2023} operates in what they call superstrong coupling regime, which arises when the atomic linewidth exceeds the cavity mode spacing. The resulting three-wave mixing interaction induces multiphoton processes across many modes, which is not our case.
Furthermore, ultrastrong coupling, as realized in our work, occurs in a different regime: when the light–matter coupling strength exceeds 10\% of the cavity photon energy. This system is described by a Generalized Dicke Hamiltonian including counter-rotating terms, leading to strong hybridization of the qubits with a cavity mode.}

%%%%%%%%%%
\begin{figure}[t!]
\centering
\includegraphics[width=85mm]{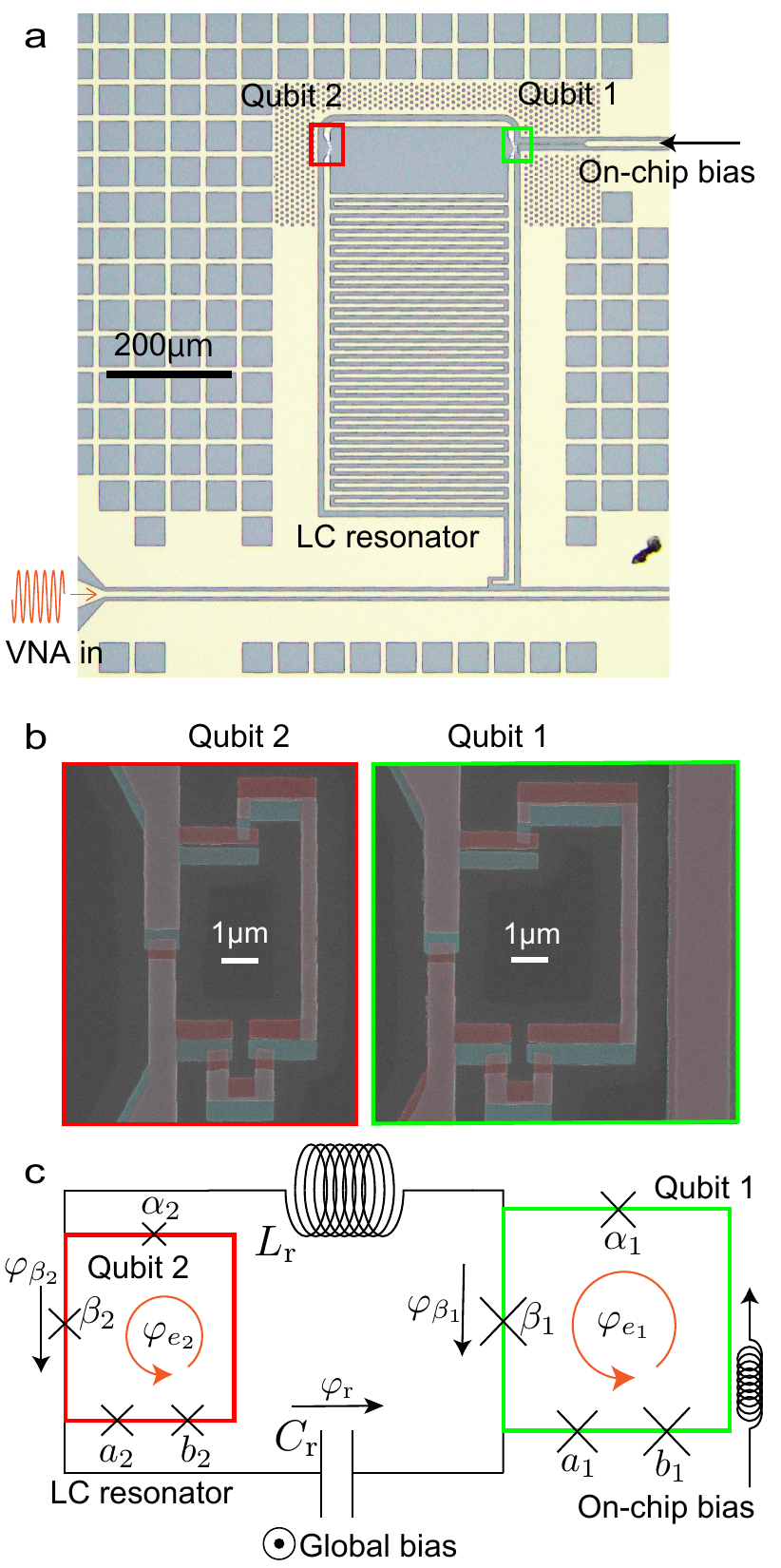}
    \caption{\textbf{Device. a}
    Optical microscopy image of the measured sample. The sample holder has a coil to bias a uniform magnetic field from the back surface of the chip. Qubit 1 has a local bias line that changes the magnetic flux of the qubit loop. The spectrum is measured using a vector network analyzer (VNA) for probing and reading from the transmission line shown below the circuit.
    \textbf{b} False-color SEM images of qubits 1 and 2. The design parameters of both qubit junctions are the same.
    \add{Different colors represent different layers of aluminum deposited via double-angle shadow evaporation.}
    \textbf{c} Equivalent circuit diagram of the sample. 
    \add{
    The symbols $\alpha_i$, $\beta_i$, $a_i$, and $b_i$ ($i \in \{1,2\}$) label each Josephson junction, while $\varphi$ denotes the phase difference across a circuit component.}
    }
\label{Fig1}
\end{figure}

%%%%%%%%%%%%%%%%%%%%%%
\begin{figure*}[t!]
\centering
\includegraphics[width=180mm]{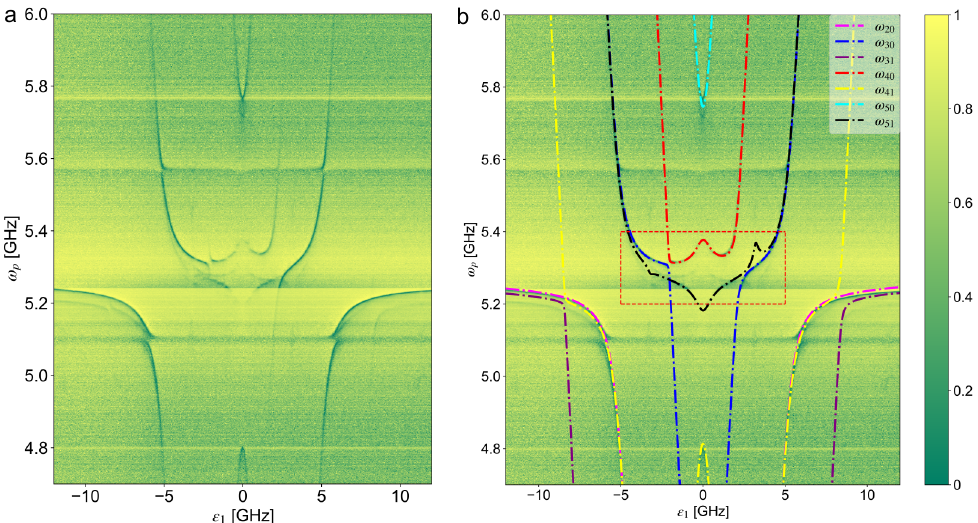}
    \caption{\textbf{Transmission spectra.} Pump frequency $\omega_p$ from the vector network analyzer versus the persistent current energy $\varepsilon_1$ of qubit 1. 
    \textbf{a} Raw data of the observed single-tone spectrum of the sample shown in Fig.~\ref{Fig1}. 
    \textbf{b} Observed single-tone spectrum with fitted curves corresponding to the state transition frequencies $\omega_{ij}$ between the $i$-th and $j$-th eigenstates of Hamiltonian~\eqref{HRabi}.
    The fit parameters are $g_1/2\pi=3.33$,  $g_2/2\pi=3.45$,  $\Delta_\mathrm{1}/2\pi=1.31$,  $\Delta_\mathrm{2}/2\pi=1.27$, $\omega_\mathrm{r}/2\pi=5.15$, and $\varepsilon_2/2\pi=-3.22$~GHz.
    At around $\omega_p/2\pi=5.09$~GHz and 5.57~GHz, parasitic modes can be seen, which originate from, for example, sample ground planes and/or the measurement environment, which includes the sample holder and microwave components coupled to the system.}
\label{Fig2}
\end{figure*}
%\section*{Results}
% \section*{Device}
\section*{Results}
\subsection*{Device}

Figure~\ref{Fig1}a shows an optical microscopy image of the artificial-atom--resonator circuit. The $LC$ resonator is composed of an interdigital capacitor and a line inductor made of a superconducting thin film~\cite{miyanaga_ultrastrong_2021,zotova_compact_2023}.
The two flux qubits are inductively coupled to the $LC$ resonator via a Josephson junction (Fig.~\ref{Fig1}b), which increases the strength of couplings to the ultrastrong regime.
Small dots around the two qubits are flux traps that prevent vortex fluctuations during the measurements~\cite{kroll_magnetic-field-resilient_2019}.
The energies of the flux qubits~\cite{chiorescu_coherent_2004} can be changed by applying an external magnetic flux to the loop from a global coil and using an on-chip bias line.
Figure~\ref{Fig1}c shows the equivalent circuit with lumped elements and Josephson junctions.

The Hamiltonian of the entire system is~\cite{tomonaga_quasiparticle_2021,billangeon_circuit-qed-based_2015,PhysRevB.73.174526}
\begin{align}
\hat{\mathcal{H}}_{\mathrm{tot}}=\hat{\mathcal{H}}_\mathrm{q1}+\hat{\mathcal{H}}_\mathrm{q2}+\hat{\mathcal{H}}_\mathrm{r}+\hat{\mathcal{H}}_{\mathrm{int}} 
\,,
\label{eq:TotalHami}
\end{align}
where $\hat{\mathcal{H}}_{\mathrm{q}k}$ ($k=1,2$), $\hat{\mathcal{H}}_\mathrm{r}$, and $\hat{\mathcal{H}}_{\mathrm{int}}$ represent the qubits, resonator, and atom--resonator plus atom--atom couplings, respectively.
The Hamiltonian of the resonator is $\hat{\mathcal{H}}_\mathrm{r}=\hbar\omega_\mathrm{r}\qty(\hat{a}^\dagger \hat{a}+1/2)$,
where $\omega_\mathrm{r}\equiv 1/\sqrt{L_\mathrm{r}C_\mathrm{r}}$ is the resonance frequency, $\hat{a}\equiv(\hat{\phi}_\mathrm{r}-iZ_\mathrm{r} \hat{q}_\mathrm{r})/\sqrt{2\hbar Z_\mathrm{r}}$ is the annihilation operator, $Z_\mathrm{r}=\sqrt{L_\mathrm{r}/C_\mathrm{r}}$ is the characteristic impedance of the $LC$ resonator, and $\hat{q}_\mathrm{r}$ is the conjugate variable of $\hat{\phi}_\mathrm{r}=\Phi_0 \hat{\varphi}_\mathrm{r}$.
Here, $\Phi_0$ is the flux quantum and the flux {${\varphi}_\mathrm{r}$ is defined in Fig.~\ref{Fig1}c}.
The Hamiltonian of the $k$-th artificial atom is
\begin{align}
\hat{\mathcal{H}}_{{\mathrm{q}k}}\equiv4E_{\mathrm{c}k}\hat{\vb{q}}_k^\mathrm{T} \vb{M}_k^{-1}{\hat{\vb{q}}_k}
+E_\mathrm{Lr}\hat{\varphi}_{\beta k}^2
+\hat{\mathcal{U}}_{\mathrm{J}k}\,,
\label{eq:Qhami}
\end{align}
where $E_{\mathrm{c}k}$ is the charging energy of the Josephson junction $a_k$ (see Fig.~\ref{Fig1}b and \ref{Fig1}c), $\varphi_{\beta k}$ represents the phase differences in each $\beta$-junction of qubits, $\vb{M}_k$ is the normalized mass matrix, $E_\mathrm{Lr}=\Phi_0^2/(2L_\mathrm{r})$, and $\hat{\mathcal{U}}_{\mathrm{J}k}$ is the qubit potential energy of Josephson junctions:
\begin{align}
        \!\!\! \hat{\mathcal{U}}_{\mathrm{J}k}(\hat{\varphi}_{\mathrm{e}k}) =&\! -\!E_{\mathrm{J}k}
        \!\left[\,
            \beta_k\cos{(\hat{\varphi}_{\beta k})}
            \!+\!\cos{(\hat{\varphi}_{ak})}\!+\!\cos{(\hat{\varphi}_{bk})}
        \right. \notag \\ 
        &\hspace{2.5em} \left.
        +\alpha_k\cos{(\varphi_{\mathrm{e}k}-\hat{\varphi}_{ak}-\hat{\varphi}_{bk}-\hat{\varphi}_{\beta k})}
        \right] \,.
    \label{eq:uj}
\end{align}
Here, $E_{\mathrm{J}k}$ is the current energy of the Josephson junction $a_k$, $\vp_{ik}$ ($i=\alpha, a, b$) are the phase differences in each junction $\alpha_k, a_k$, and $b_k$, and $\varphi_{\mathrm{e}k}$ represents the external flux for the loop of each atom. The interaction Hamiltonian
\begin{align}
\hat{\mathcal{H}}_{\mathrm{int}}=-E_\mathrm{Lr}(\hat{\varphi}_{\beta1}\hat{\varphi}_\mathrm{r}-\hat{\varphi}_{\beta2}\hat{\varphi}_\mathrm{r}+\hat{\varphi}_{\beta1}\hat{\varphi}_{\beta2})
    \label{eq:Hint}
\end{align}
is obtained from the boundary condition (Kirchhoff's voltage law) of the loop forming the resonator with the elements $L_\mathrm{r}$ and $C_\mathrm{r}$.

By approximating each atom as a two-level system~\cite{yoshihara_hamiltonian_2022} on the basis of persistent currents of the superconducting loop, we obtain the total Hamiltonian in Eq.~\eqref{eq:TotalHami} as
\begin{align}
\!\!\! 
\hat{\mathcal{H}}_\mathrm{tot}/\hbar \simeq & \; \omega_\mathrm{r}\hat{a}^\dagger \hat{a}\!+\!\frac{\varepsilon_1}{2}\hat{\sigma}_{z1}\!+\!\frac{\Delta_1}{2}\hat{\sigma}_{x1}\!+\!\frac{\varepsilon_2}{2}\hat{\sigma}_{z2}\!+\!\frac{\Delta_2}{2}\hat{\sigma}_{x2} \notag
\\&-(g_1\hat{\sigma}_{z1}-g_2\hat{\sigma}_{z2})(\hat{a}^\dagger+\hat{a}) 
-\frac{2g_1 g_2}{\omega_\mathrm{r}}\hat{\sigma}_{z1}\hat{\sigma}_{z2}\,,
\label{HRabi_z}
\end{align}
where $\varepsilon_k$ is the persistent current energy of each qubit, $\Delta_k$ is the qubit energy gap when $\varepsilon_k=0$, while $\hat{\sigma}_{zk}$ and $\hat{\sigma}_{xk}$ are the Pauli matrices for the $k$-th qubit. We define $\varepsilon_k>0$ when the qubit current flows anticlockwise and vice versa.

After a unitary transformation that diagonalizes the atomic Hamiltonians $\hat{\mathcal{H}}_{\mathrm{qk}}$, we obtain a generalized Dicke Hamiltonian~\cite{jaako2016ultrastrong} with spin-spin interaction:
\begin{align}
\hat{\mathcal{H}}_\mathrm{tot}/\hbar \simeq & \;
\omega_\mathrm{r}\hat{a}^\dagger \hat{a} 
+ \frac{\omega_\mathrm{q1}}{2}\hat{\sigma}_{z1} 
+ \frac{\omega_\mathrm{q2}}{2}\hat{\sigma}_{z2} \notag \\
&
- (g_1\hat{\Lambda}_1-g_2\hat{\Lambda}_2)(\hat{a}^\dagger+\hat{a}) 
-\frac{2g_1 g_2}{\omega_\mathrm{r}}\hat{\Lambda}_1\hat{\Lambda}_2\,,
\label{HRabi}
\end{align}
where $\omega_{\mathrm{q}k}={\rm sgn}(\varepsilon_k)(\varepsilon_k^2+\Delta_k^2)^{1/2}$ is the qubit frequency and $\hat{\Lambda}_k=(\cos{\theta_k}\,\hat{\sigma}_{xk}+\sin{\theta_k}\,\hat{\sigma}_{zk})$ gives the direction of the interaction, with $\theta_k\simeq-\arctan(\varepsilon_k/\Delta_k)$ \add{(see Methods for more details)}. 
For $\theta_k=0$ $(\varepsilon_k=0)$, the interaction is purely transverse. When $\theta_k\neq0$, the interaction has a longitudinal component and the one--photon--exciting--two--atoms effect is allowed.

%%%%%%%%%%
\begin{figure*}[t!]
\centering
\includegraphics[width=180mm]{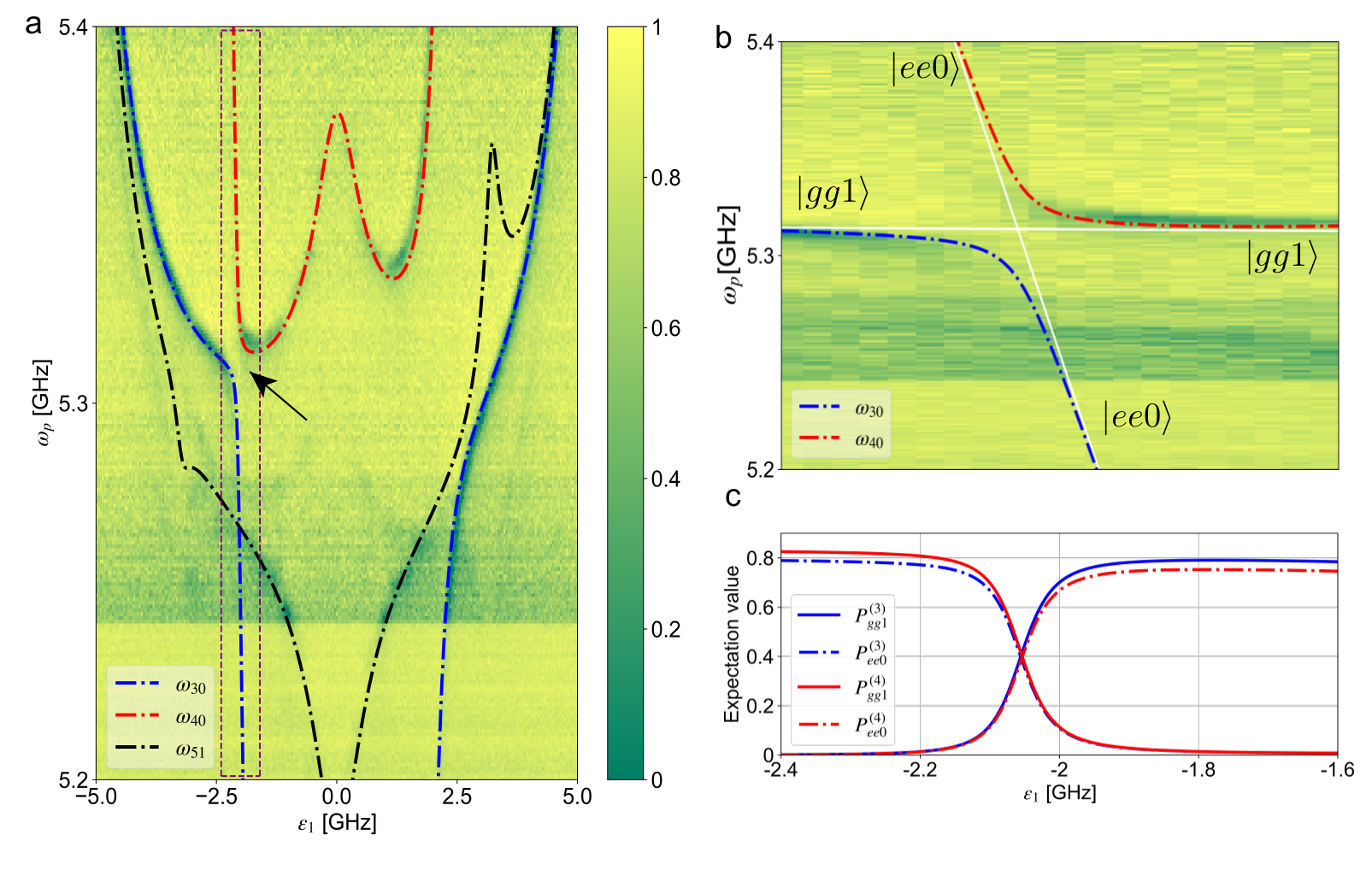}
    \caption{
    \textbf{Anticross between $\ket{gg1}$ and $\ket{ee0}$. a} Enlarged view of the central part of the spectrum in Fig.~\ref{Fig2}b with fitting curve. The fitting reproduces the spectrum well.
    \textbf{b} Enlarged image of the anticrossing between $\omega_{30}$ and $\omega_{40}$.
    The white lines represent the eigenmodes of $\ket{gg1}$ and $\ket{ee0}$ in the non-interacting Hamiltonian \add{(see Methods for more details)}.
    \textbf{c} Projection of the third and fourth eigenstates calculated using Hamiltonian in Eq.~\eqref{HRabi} with the fitting parameters to the bare states $\ket{gg1}$ and $\ket{ee0}$. 
    }
\label{Fig3}
\end{figure*}
%%%%%%%%%%%%%%%%%%%%%%

%%%%%%%%%%%%%%%%%%%%%%

\subsection*{Energy spectrum}

Figure~\ref{Fig2}a shows the raw data of the measured spectrum as a function of the persistent current energy $\varepsilon_1$ of qubit 1, which are obtained after fixing the value of $\varepsilon_2/2\pi$ at $-3.22$~GHz when $\varepsilon_1=0$.
In Fig.~\ref{Fig2}b, the spectrum is fitted with the numerically calculated transition frequencies $\omega_{ij}$ between the $i$-th and $j$-th eigenstates of the total Hamiltonian $\hat{\mathcal{H}}_\mathrm{tot}$.
The persistent current energy for qubit 2 and the resonator frequency are affected by the external magnetic flux applied to qubit 1~\cite{yoshihara_superconducting_2017}.
Thus, to derive the transition frequencies $\omega_{ij}$, we substitute $\varepsilon_2\xrightarrow{}\varepsilon_{2}+A\varepsilon_1$ and $\omega_\mathrm{r}\xrightarrow{}\omega_\mathrm{r}(1+B_\pm\varepsilon_1)$ in Eq.~\eqref{HRabi_z}, where $A$ and $B_\pm$ are small fitting parameters listed in Table~\ref{table:SampleData}.
We use two different values for $B_\pm$ because the spectrum is asymmetric with respect to the sign of $\varepsilon_1$, i.e., $B_+$ is used when $\varepsilon_1\ge0$ and vice versa.
Including $A$ and $B_{\pm}$, we use 11 parameters in total for the fit. These also include the bias current offset $I_{b0}$, when $\varepsilon_1=0$, and the persistent current coefficient $\tilde{\varepsilon}_0$ to derive $\hbar\varepsilon_1=I_\mathrm{p}\Phi_0(\varphi_\mathrm{e1}-0.5)=\hbar\tilde{\varepsilon}_0(I_\mathrm{b}-I_{b0})$, where $I_\mathrm{p}$ is the persistent current of qubit 1, and $I_b$ is the bias current from the room-temperature current source.
We use a photo-processing technique to obtain peak points from the spectrum~\cite{sato_three-dimensional_1998,walt_scikit-image_2014,tomonaga_quasiparticle_2021} and the quantum toolbox in Python (QuTip) for numerical calculations~\cite{johansson_qutip_2012,johansson_qutip_2013}.

%%%%%%%%%%%%%%%%%%%%%%%%%%%%%
\begin{table}[t]
    \caption{List of the fitting parameters used in Figs.~\ref{Fig2} and~\ref{Fig3}.
    We use the first eight listed parameters for an initial fit. Afterwards, we add the last three parameters to slightly modify the fit. The initial parameters $\omega_r$, $\tilde{\varepsilon}_0$, and $I_{b0}$ can be obtained from the spectrum. Also, the qubit parameters $g_{i}$ and $\Delta_{i}$ can be estimated from the design parameters.
    }
    \centering
    \label{table:SampleData}
  \begin{tabularx}{0.48\textwidth}{Xccc}
    \hline
    Name        & \quad  Symbol \quad\quad & Value \quad   &  \, Unit \quad\\ 
    \hline \hline
Resonator frequency  & $\omega_r/2\pi$        & 5.15   &  $\si{\giga\hertz}$  \\
Coupling constant of $Q_1$        & $g_1/2\pi$        &  3.33  &  $\si{\giga\hertz}$  \\
Coupling constant of $Q_2$      & $g_2/2\pi$          &  3.45  &  $\si{\giga\hertz}$    \\
Energy gap of $Q_1$     & $\Delta_1/2\pi$          &  1.31  &  $\si{\giga\hertz}$    \\
Energy gap of $Q_2$           & $\Delta_2/2\pi$  &  1.27  &  $\si{\giga\hertz}$   \\
Persistent\,current\,energy\,of\,$Q_2$   & $\varepsilon_2/2\pi$  &  -3.22  &  $\si{\giga\hertz}$    \\
Persist current coefficient   & $\tilde{\varepsilon}_0/2\pi$   &  201.6 &  $\si{\giga\hertz/mA}$    \\
Bias current offset  & $I_{b0}$   &  0.547  &  $\mathrm{mA}$    \\
Crosstalk coefficient & $A$         & -9.43  &  $\times10^{-3}$   \\
Res.\,freq.\,modification $\varepsilon_1\!>\!0$  & $B_+$         &  $0.78$  &  $\times10^{-3}$    \\
Res.\,freq.\,modification $\varepsilon_1\!<\!0$ & $B_-$  &  $0.73$  &  $\times10^{-3}$    \\
    \hline
\end{tabularx}
\end{table}
%%%%%%%%%%%%%%%%%%%%%%%%%%%

Flux qubits 1 and 2 are almost identical except for the loop size; consequently, they have similar fitting parameters; i.e., $\Delta_\mathrm{q1}\simeq\Delta_\mathrm{q2}\simeq0.25\,\omega_\mathrm{r}$.
We find atom-resonator coupling rates of $g_{1}/\omega_\mathrm{r}=0.67$ and $g_{2}/\omega_\mathrm{r}=0.69$, indicating that the artificial atoms are ultrastrongly coupled with the resonator.

\subsection*{One photon simultaneously excites two atoms}

We indicate with $\ket{\psi_i}$ the eigenstate of the system Hamiltonian $\hat{\mathcal{H}}_\mathrm{tot}$ with eigenenergies $\hbar \omega_{i0}$.
The $\omega_{\mathrm{q}i}\hat\sigma_{zi}/2$ terms in Eq.~(\ref{HRabi}) define the ground $\ket{g}$ and excited $\ket{e}$ atomic bare states.

In Fig.~\ref{Fig3}a, which is an enlarged view of the red dashed rectangle in Fig.~\ref{Fig2}b, the black arrow indicates the anticrossing between the eigenstates $\ket{\psi_3}$ and $\ket{\psi_4}$ (see Fig.~\ref{Fig3}b), with eigenfrequencies $\omega_{30}$ and $\omega_{40}$. 
In agreement with this anticrossing, Fig.~\ref{Fig3}c shows the numerically calculated projection $P_{j}^{(i)}\equiv\abs{\braket{\psi_i}{j}}^2$ of the third and fourth eigenstates $\ket{\psi_i}$ $(i=3,4)$ on the bare states $\ket{j}=\{\ket{gg1},\ket{ee0}\}$ as a function of $\varepsilon_1$.
Here, it is possible to see that the third and fourth eigenstates are the approximate symmetric and antisymmetric superpositions of $\ket{gg1}$ and $\ket{ee0}$, respectively. Considering also that the sum of the dressed qubit frequencies is nearly equal to the dressed resonator frequency, the anticrossing is the signature of the one--photon--exciting--two--atoms effect ~\add{(see Methods for more details).}
Half of the minimum split between $\omega_{30}$ and $\omega_{40}$ in the spectrum gives the effective coupling between $\ket{gg1}$ and $\ket{ee0}$, that is 22.8~MHz \add{(see Supplementary Information)}.

With respect to the theoretical prediction in ref.~\onlinecite{garziano_one_2016} ($g/\omega_\mathrm{r}\simeq0.1$--$0.2$), our system has a much larger coupling ($g/\omega_\mathrm{r}\simeq0.7$). This implies that the system eigenstates should have a strong dressing, and in principle we could not observe a clean ``one--photon--exciting--two--atoms'' effect. 
On the contrary, Fig.~\ref{Fig3}c shows that the dressing is low for those states, and, as shown in Figs.~\ref{FigS0}, our system can still be considered formed by separated two two-level atoms and one cavity mode with shifted eigenfrequencies.
This behavior is heuristically justified by the fact that spin--spin and spins--light couplings are competing interactions and that the longitudinal interaction ``decouples'' for specific values of the signs of $\varepsilon_1$ and $\varepsilon_2$. The signature of this ``longitudinal decoupling'' is given by the asymmetry in the spectrum with respect to the sign of $\varepsilon_1$.
Assuming that there are only longitudinal couplings, in the interaction part of Eq.~(\ref{HRabi}), the operator $(\hat{a}^\dagger+\hat{a})$ should generate coherent states of light in the ultrastrong coupling regime \cite{stassi2018long}. However, considering $\varepsilon_1<0$ and $\varepsilon_2<0$, the atomic states are not associated with coherent states if $M=m_1-m_2=0$, where $m_k=\pm 1$ is the eigenstate of $\hat{\sigma}_{zk}$ $(k=1,2)$ ~\add{(see Methods for more details)}. As a result, the ground $\ket{gg}$ and excited $\ket{ee}$ states, which have $M=0$, have no coherent states associated with them. 
Nevertheless, the transverse interactions still affect our system generating a small dressing that reduces the projections $P_{gg1}^{(4)}$ and $P_{ee0}^{(3)}$ to almost $0.8$ at $\varepsilon_1/2\pi=-2.4$~GHz.

\section*{Discussion}

We measured the spectrum of a circuit composed of two artificial atoms ultrastrongly coupled to an $LC$ resonator. The generalized Dicke Hamiltonian with spin--spin interaction correctly describes the measured spectrum.
At the energy where the sum of the atomic energies almost matches that of the resonator, we observed one anticrossing between the states $\ket{gg1}$ and $\ket{ee0}$.
This experimentally confirms the recent theoretical prediction that one photon can simultaneously excite two atoms~\cite{garziano_one_2016}, opening a new chapter in quantum nonlinear optics.

Clearly, only the avoided-level crossing and not the excitation process itself has been experimentally observed, because our \erase{sample}\add{experiment} was not designed \erase{for the}\add{to observe} Rabi oscillation\add{s}\erase{ experiment}. However, our \add{spectroscopic} observations prove the existence of this excitation.

Future work will involve reading out the qubit and photon states~\cite{felicetti_parity-dependent_2015} as well as observing the one--photon--exciting--two--atoms dynamics. Theoretically, the reduction of dressing and the asymmetry in the spectrum could be further investigated.
These studies could also be extended to explore, for example, photon down- and up-conversions~\cite{inomata_microwave_2014,Kockum_Deterministic} and ultrafast two qubit gates~\cite{romero_ultrafast_2012,wang_ultrafast_2017}.

\section*{Methods}
\subsection*{Circuit Hamiltonian}
Here, we describe the circuit Hamiltonian calculation in detail.
The branch fluxes across the circuit elements, which are Josephson junctions, the inductance $L_\mathrm{r}$, and the capacitance $C_\mathrm{r}$, follow Kirchhoff's voltage laws:
\begin{align}
    &\varphi_{\beta1}+\varphi_{\alpha1}+\varphi_{a1}+\varphi_{b1} =\varphi_{\mathrm{e}1}\,, \\
    &\varphi_{\beta2}+\varphi_{\alpha2}+\varphi_{a2}+\varphi_{b2} =\varphi_{\mathrm{e}2}\,, \\
    &\varphi_\mathrm{cr}+\varphi_\mathrm{\ell r}+\varphi_{\beta1}-\varphi_{\beta2}=0\,, 
    \label{eq:KVL}
\end{align}
where $\varphi_\mathrm{cr}$ and $\varphi_\mathrm{\ell r}$ represent fluxes between the resonator capacitor and the inductor.
The total Lagrangian of the circuit is
\begin{align}
    \mathcal{L}_\mathrm{tot}=\mathcal{K}_\mathrm{J1}+\mathcal{K}_\mathrm{J2}
    -\mathcal{U}_\mathrm{J1}-\mathcal{U}_\mathrm{J2}+\mathcal{L}_\mathrm{r}\,,
    \label{eq:Ltot}
\end{align}
where
\begin{align}
    \mathcal{K}_{\mathrm{J}k}
    &=\frac{\cj}{2}\left[\,
    \beta\dot{\phi}_{\beta k}^2
    +\dot{\phi}_{ak}^2+\dot{\phi}_{bk}^2 \right. \notag \\ 
    &\quad\quad\quad\quad\left.
    +\alpha_i(\dot{\phi}_{\beta k}+\dot{\phi}_{ak}+\dot{\phi}_{bk})^2\right] \,,
\end{align}
\begin{align}
    \begin{split}
        \mathcal{U}_{\mathrm{J}k}=-\ej&
        \left[
        \beta_i\cos(\varphi_{\beta k})
        +\cos{(\varphi_{ak})} +\cos{(\varphi_{bk})} 
        \right.  \\ 
        &\left.
        +\alpha_i\cos{(\varphi_{\mathrm{e}k}-\varphi_{\beta k}-\varphi_{ak}-\varphi_{bk})}
        \right]
        \,,
    \end{split}
    \label{eq:uj}
\end{align}
\begin{align}
    \mathcal{L}_\mathrm{r}
    =\frac{C_\mathrm{r}}{2}\dot{\phi}_{\mathrm{cr}}^2-\frac{1}{2L_\mathrm{r}}(\phi_\mathrm{cr}+\phi_{\beta 1}-\phi_{\beta 2})^2\,.
    \label{Lr}
\end{align}
Here, $k\in \{1,2\}$ indicates the qubit, and the sub-index cr indicates the capacitor of the resonator.
The qubit kinetic energy part of the Lagrangian in Eq.~\eqref{eq:Ltot} becomes
\begin{align}
    \mathcal{K}_{\mathrm{J}i}= \frac{1}{2} \dot{\boldsymbol{\upphi}}{\!} ^\mathrm{T}\vb{M}\dot{\boldsymbol{\upphi}} \,,
    \label{eq:Kq}
\end{align}
where $\boldsymbol{\upphi}_k \equiv \mqty(\phi_{\beta k} & \phi_{ak} & \phi_{bk})^\mathrm{T}$ and the mass matrix is given by
\begin{align}
    \vb{M}_k &= C_\mathrm{J}\mqty (
        \beta_k+\alpha_k & \alpha_k & \alpha_k  \\
        \alpha_k & 1+\alpha_k & \alpha_k   \\
        \alpha_k & \alpha_k & 1+\alpha_k \\
        ) \,.
        \label{eq:matrix}
\end{align}

Using the canonical conjugate $q_i= \partial\mathcal{}{L}_\mathrm{tot} / \partial \dot{\phi}_i$ for $\dot{\phi}_i$, where $i\in\{\beta_k,a_k,b_k\}$, we can rewrite Eq.~\eqref{eq:Kq} as
\begin{align}
    \mathcal{K}_{\mathrm{J}k}=\frac{1}{2}\vb{q}_k^\mathrm{T}\vb{M}^{-1}\vb{q}_k\,.
\end{align}
Then, we obtain the total Hamiltonian of the circuit as
\begin{align}
   \mathcal{H}_{\mathrm{tot}}&= \sum_k\vb{q}_k^\mathrm{T} \dot{\boldsymbol{\upphi}}_k-\mathcal{L}_{\mathrm{tot}} \notag\\
    &=\sum_k (4E_\mathrm{c}\tilde{\vb{q}}_k^\mathrm{T} \tilde{\vb{M}}_k^{-1}\tilde{\vb{q}}_k+E_\mathrm{Lr}\varphi_{\beta k}^2
    +\mathcal{U}_{\mathrm{J}k}) \notag\\
    &\quad\,\,\,+\mathcal{H}_\mathrm{r}+\mathcal{H}_\mathrm{int} \,,
    \label{eq:TotalHami_A}
\end{align}
where $2e\tilde{q}=q$ and $C_\mathrm{J}\tilde{\vb{M}}=\vb{M}$. The term $E_\mathrm{Lr}\varphi_{\beta k}$ originates from Eq.~\eqref{Lr} since we define
\begin{align}
  \mathcal{H}_\mathrm{r}\equiv\frac{C_\mathrm{r}}{2}\dot{\phi}_{\mathrm{cr}}^2+\frac{1}{2L_\mathrm{r}}\phi_\mathrm{cr}^2\,
\end{align}
as a bare $LC$ resonator.
We now replace the canonical values with the operators $\vp\rightarrow\hat{\vp}$ and $q\rightarrow\hat{q}$ and impose their relation $[\hat{\vp}, \hat{q}]=i$. For the resonator, we introduce the creation and annihilation operators. Thereafter,
we expand the quantized total Hamiltonian using the eigenvectors $\ket{i}_i$ (${i}\in \mathbb{N}$) of the atom Hamiltonians ($\hat{\mathcal{H}}_\mathrm{tot}-\hat{\mathcal{H}}_\mathrm{r}-\hat{\mathcal{H}}_\mathrm{int}$),
\begin{align}
\!\!\!\hat{\mathcal{H}}_{\mathrm{tot}}=&\hbar\sum_{i}{(\Omega_{i}^{(1)}\ket{i}_1\bra{i}_1+\Omega_{i}^{(2)}\ket{i}_2\bra{i}_2)}
+\hat{\mathcal{H}}_\mathrm{r} \notag \\ \notag
&\!\!-\hbar\sum_{i,j}{(g_{ij}^{(1)}\ket{i}_1\bra{j}_1\!\!-g_{ij}^{(2)}\ket{i}_2\bra{j}_2)(\hat{a}^\dagger\!+\hat{a})}    \\
&-E_\mathrm{L}\sum_{i,j}{g_{ij}^{(1)}g_{ij}^{(2)}\ket{i}_1\ket{i}_2\bra{j}_1\bra{j}_2}\,,
\label{Hket}
\end{align}
where $\hbar\Omega_{i}^{(k)}$ is the $i$-th eigenenergy of atom $k$ and $\hbar g_{ij}^{(k)}=I_\mathrm{zpf}\Phi_0\mel{i}{\hat{\varphi}_{\beta k}}{j}$ is the coupling matrix element ($I_{\mathrm{zpf}}=\sqrt{\hbar\omega_\mathrm{r}/2L_\mathrm{r}}$).
After truncating the higher state of the flux qubits, we obtain the Hamiltonian of two two-level atoms and a resonator, Eqs.~\eqref{HRabi_z} and~\eqref{HRabi}.
In Eqs.~\eqref{HRabi_z} and~\eqref{HRabi}, the spin-spin interaction reduces the current flowing in the resonator loop, which in this system is the ferromagnetic coupling.

%%%%%%%%%%%%%%%%%%%%%%%%%%%%%

\begin{figure}[]
\centering
\includegraphics[width=85mm]{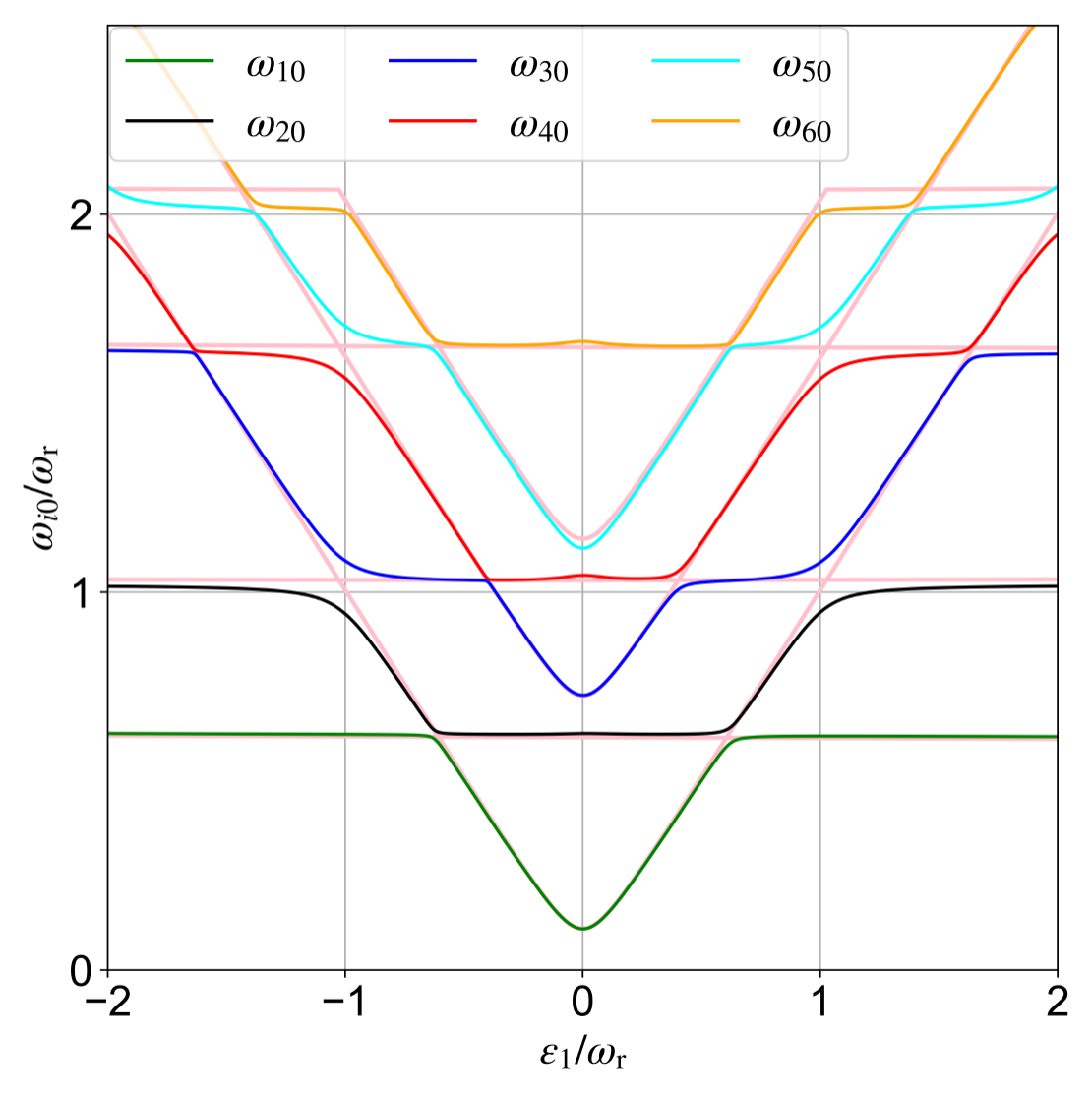}
\caption{
Comparison between the lowest eigenstates of the Hamiltonian in Eq.~\eqref{HRabi} and the non-interacting Hamiltonian (light pink curves). The non-interacting Hamiltonian is derived from Eq.~\eqref{HRabi} with $g_1=g_2=0$, shifting the atom frequencies. The shifts are chosen such that $\Delta_{1,2}$ $\rightarrow$ $\Delta_{1,2}\exp{-2(g_{1,2}/\omega_\mathrm{r})^2}$. For simplicity, we ignore the qubit bias current crosstalk, A= 0.
}
\label{FigS1}
\end{figure}
%%%%%%%%%%%%%%%%%%%%%%%%%%%%%%%%%%%%%%%%%
%%%%%%%%%%%%%%%%%%%%%%%%%%%%%%%
\begin{figure}[h!]
\centering
\includegraphics[width=85mm]{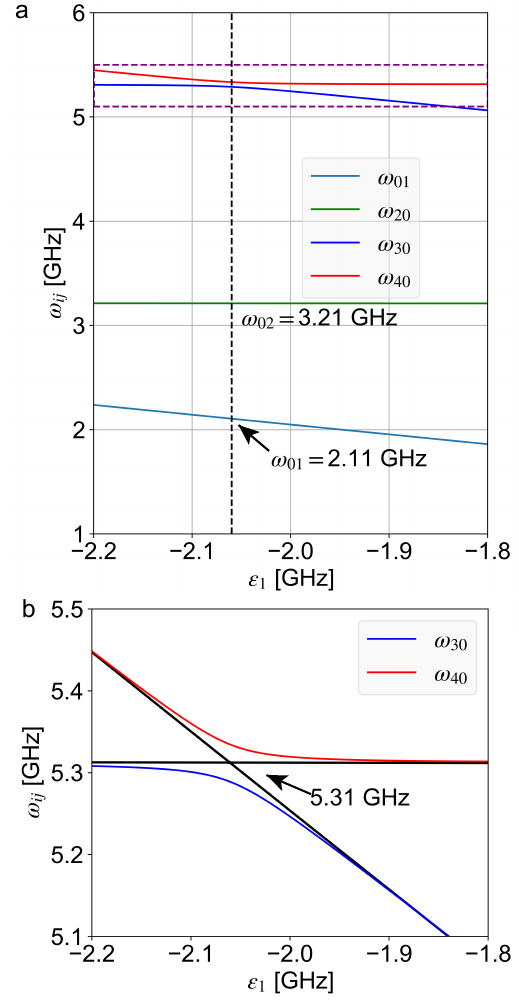}
\caption{\textbf{a} Numerically calculated spectrum as a function of $\varepsilon_1$ around the anticross point between $\omega_{30}$ and $\omega_{40}~~(\varepsilon_1<0)$. \textbf{b} Enlarged image of the purple dashed square in \textbf{a}. Black lines show $\omega_{30}$ and $\omega_{40}$ derived from the no-interacting Hamiltonian.}

\label{FigS0}
\end{figure}
%%%%%%%%%%%%%%%%%%%%%%%%%%%%
\subsection*{One--photon two--atoms energy exchange}
Figure~\ref{FigS1} shows numerically calculated transition frequencies of the system using the fit parameters over a wider range. For comparison, we overlaid the eigenvalues of the Hamiltonian in Eq.~\eqref{HRabi} with the eigenvalues calculated when $g_1=g_2=0$. In the latter case, we shifted the atoms and cavity frequencies to consider the dressing effect. From Fig.~\ref{FigS1}, it is possible to notice that the eigenvalues of the interacting and non-interacting ($g_1=g_2=0$) Hamiltonians almost coincide, proving that our system, operating in the ultrastrong coupling regime, can still be considered composed of two independent atoms and a resonator that interact in the crossing points.

Figure~\ref{FigS0}a {is an enlarged image of Fig.~\ref{FigS1}, with the range around the anticrossing between $\omega_{30}$ and $\omega_{40}~~(\varepsilon_1<0)$.
In our circuit, $\omega_{01}$ and $\omega_{02}$ are the frequencies of dressed qubits 1 and 2, respectively.
From the enlarged image of the anticrossing in Fig.~\ref{FigS0}a and ~\ref{FigS0}b, it is possible to see that the transition frequency of $\ket{gg1}$ at the anticrossing point is $5.312~\si{\giga\hertz}$, which is close to the sum of the frequencies of the dressed qubits, $\omega_{01}+\omega_{02}=5.318~\si{\giga\hertz}.$

% %%%%%%%%%%%%%%%%%%%%%%%%%%%%%
% \begin{figure*}[t]
% \centering
% \includegraphics[width=180mm]{FigS2.pdf}
% \caption{\textbf{a} shows the maximum values of the projections $P_{gg1}(g/\omega_\mathrm{r})$ and $P_{ee0}(g/\omega_\mathrm{r})$ at the antisplitting point. The green dotted vertical line corresponds to the value that maximizes $g_{30-40}$.
% \textbf{b} shows the effective coupling constant $g_{30-40}$ plotted against the coupling ratio $g/\omega_\mathrm{r}$.
% The red and blue stars represent $g_1/\omega_\mathrm{r}=0.64$ and $g_2/\omega_\mathrm{r}=0.67$, respectively.
% }
% \label{FigS2}
% \end{figure*}
% %%%%%%%%%%%%%%%%%%%%%%%%%%%%%

It has been shown that when an atom interacts longitudinally with light, the atomic states are associated with coherent states of light~\cite{stassi2018long}. In our system, the generation of coherent states, given by the longitudinal coupling of the atoms with the resonator, depends on the signs of $\varepsilon_1$ and $\varepsilon_2$. In the following derivation, we show that when $\varepsilon_1<0$ and $\varepsilon_2<0$, the atomic states $\ket{ge}$ and $\ket{eg}$ are associated with the coherent states of light, while states $\ket{gg}$ and $\ket{ee}$ are not associated with those.

Considering the system Hamiltonian [Eq.~\eqref{HRabi}] with $\Delta_k=0$, $g_1=g_2=g$, and substituting $\sigma_{zk}$ with its eigenvalue $m_k=\pm1$, we can write the following:
\begin{align}
   \hat{\mathcal{H}}_{s}/\hbar=\frac{\vert\varepsilon_1\vert}{2}m_1+\frac{\vert\varepsilon_2\vert}{2}m_2+\omega_\mathrm{r}\hat{a}^\dagger \hat{a}+gM(\hat{a}^\dagger+\hat{a})\,,
   \label{Hsi}
\end{align}
with $M={\rm sgn}(\varepsilon_2)m_2-{\rm sgn}(\varepsilon_1)m_1$. Performing the substitution  $\hat{a}=\hat{b}-Mg/\omega_\mathrm{r}$ , we obtain
\begin{equation}
   \hat{\mathcal{H}}_{s}/\hbar=\omega_\mathrm{r}\hat{b}^\dagger \hat{b}+\frac{\vert\varepsilon_1\vert}{2}m_1+\frac{\vert\varepsilon_2\vert}{2}m_2-M^2\frac{g^2}{\omega_\mathrm{r}}\,,
   \label{Hs}
\end{equation}
which is the Hamiltonian of a harmonic oscillator.
By applying the annihilation operator $\hat{b}$ to its ground state $\ket{0}_M$ (i.e., $\hat{b}\ket{0}_M=0$), we have
\begin{align}
   \hat{a}\ket{0}_M=-M\frac{g}{\omega_\mathrm{r}}\ket{0}_M\,.
   \label{aM}
\end{align}
From Eq.~\ref{aM}, we see that atomic states with $M=0$ are associated with the zero-photon state; while atomic states with $M=\pm2$ are associated with photonic coherent states $\ket{\pm\alpha}$. In turn, $M$ depends on the signs of $\varepsilon_1$ and $\varepsilon_2$. If $\varepsilon_1<0$ and $\varepsilon_2<0$, then $M_-\equiv M=m_1-m_2$. If $\varepsilon_1>0$ and $\varepsilon_2<0$, then $M_+\equiv M=-(m_1+m_2)$. 
Table \ref{t1} shows the eight possible states as a function of the sign of $\varepsilon_1$ when the interaction is longitudinal. This explains the asymmetry of the spectra with respect to the sign of $\varepsilon_1$.

\begin{table}[h!]
    \caption{Possible values of $M$ and relative states. We consider $\varepsilon_2<0$.}
    \centering
   % \label{table:M}
  \begin{tabularx}{0.48\textwidth}{ccccc}
  \\
     $m_1$     &  \,\,-1 &  \,\,+1  & \,\,-1 &  \,\,+1  \\\\
        \hline\\
     $m_2$     &  \,\,-1 &  \,\,-1  & \,\,+1 &  \,\,+1  \\\\
        \hline\\
     $M_-$       &  \,\,0 &  \,\,+2  & \,\,-2 &  \,\,0   \\\\
        \hline\\
        States if $\varepsilon_1<0$    &  \,\,$\ket{gg}\ket{0}$  &  \,\,$\ket{eg}\ket{-\alpha}$  &\,\,$\ket{ge}\ket{+\alpha}$ & \,\, $\ket{ee}\ket{0}$  \\\\
    \hline\\
     $M_+$       &  \,\,+2 &  \,\,0  & \,\,0 &  \,\,-2   \\\\
        \hline\\
        States if $\varepsilon_1>0$    &  \,\,$\ket{gg}\ket{+\alpha}$  &  \,\,$\ket{eg}\ket{0}$  &\,\,$\ket{ge}\ket{0}$ & \,\, $\ket{ee}\ket{-\alpha}$  \\\\
    \hline
    \label{t1}
\end{tabularx}
\end{table}

% \section*{Code availability}
% \add{All custom codes used in this study are provided as text files in the Supplementary Information.}

% \section*{Data availability}
% \add{All figures and results in this study can be reproduced using the CSV data and code provided in the supplementary information. All data are available in the main text or supplementary information, including the relevant CSV and TXT files.}


%%%%%%%%%%%%%%%%%%%%%%%%%%%%%%%%%%%%%%%%%%%

\subsection*{Acknowledgements}

We thank Y. Zhou, R. Wang, S. Shirai, S. Savasta S. Kwon, and K. Inomata for their thoughtful comments on this research.
This paper was based on results obtained from JSPS KAKENHI (Grant Number JP 22K21294,23K13048) and a project, JPNP16007, commissioned by the New Energy and Industrial Technology Development Organization (NEDO), Japan.
Supporting from JST CREST (Grant No. JPMJCR1676) and Moonshot R\&D (Grant No. JPMJMS2067) is also appreciated.
R.S. acknowledges the Army Research Office (ARO) (Grant No. W911NF1910065).
F.N. is supported in part by: the Japan Science and Technology Agency (JST) [via the CREST Quantum Frontiers program Grant No. JPMJCR24I2, the Quantum Leap Flagship Program (Q-LEAP), and the Moonshot R\&D Grant Number JPMJMS2061], and the Office of Naval Research (ONR) Global (via Grant No. N62909-23-1-2074).

\subsection*{Author contributions}
\add{A.T. designed the device, carried out the experiment, and analyzed the data. R.S. and A.T. performed theoretical and numerical calculations. H.M. carried out part of the experiment. F.N., F.Y., and J.S.T. participated in discussions and contributed to the writing and editing of the manuscript. J.S.T. also conducted the research and managed the laboratory. All authors contributed to the manuscript preparation and revision.}

\end{document}